\documentclass[aps,preprintnumbers,eqsecnum,amsmath,amssymb,showpacs,
nofootinbib]{revtex4}
\usepackage{graphicx}
\usepackage{dcolumn}
\usepackage{bm}
\usepackage{epsfig}

\begin{document} 
\title{\boldmath OPE and quark-hadron duality for two-point functions\\
of tetraquark currents in $1/N_c$ expansion}
\author{Wolfgang Lucha$^a$, Dmitri Melikhov$^{b,c,d}$, Hagop Sazdjian$^e$}
\affiliation{
$^a$Institute for High Energy Physics, Austrian Academy of Sciences,
Nikolsdorfergasse 18, A-1050 Vienna, Austria\\
$^b$D.~V.~Skobeltsyn Institute of Nuclear Physics, M.~V.~Lomonosov Moscow
State University, 119991, Moscow, Russia\\
$^c$Joint Institute for Nuclear Research, 141980 Dubna, Russia\\
$^d$Faculty of Physics, University of Vienna, Boltzmanngasse 5, A-1090
Vienna, Austria\\
$^e$Universit\'e Paris-Saclay, CNRS/IN2P3, IJCLab, 91405 Orsay, France} 
\begin{abstract}
We discuss the Operator Product Expansion (OPE) and quark-hadron duality
for two-point Green functions of tetraquark 
currents. We emphasize that the factorizable part of the OPE series for
such Green functions, including nonperturbative 
contributions described by QCD condensates, is saturated by the full
system of ordinary hadrons and therefore 
cannot have any relationship to the possible tetraquark bound states. 
Possible tetraquark bound states may be contained in nonfactorizable
parts of these Green functions. 
In the framework of the $1/N_c$ expansion in QCD$(N_c)$, nonfactorizable
parts of the two-point Green 
functions of tetraquark currents provide $N_c$-suppressed contributions
compared to the $N_c$-leading 
factorizable parts. A possible exotic tetraquark state may appear only
in $N_c$-subleading 
contributions to the QCD Green functions, in full accord with the
well-known rigorous properties of large-$N_c$ QCD. 
\end{abstract}

\pacs{11.55.Hx, 12.38.Lg, 03.65.Ge} 
\date{\today}
\maketitle

\section{Introduction
\label{Sect:I}}
The correlation functions of two local colorless currents are the simplest
gauge-invariant Green functions that have a unique decomposition in terms
of the physical hadron states. These correlation functions are defined as
vacuum expectation values of the time-ordered-products of two local
gauge-invariant quark currents taken at different locations:  
\begin{eqnarray}
\Pi(p^2)=i\int
d^4x\,e^{ipx}\left\langle 0 \left|T\{ j(x) j^\dag(0) \}
\right|0\right\rangle. 
\end{eqnarray}
We shall discuss and compare two cases: $j$ being a local bilinear
quark current and $j$ being a tetraquark current. 
The Dirac structure of the currents will be of no relevance for our
arguments 
and will not be specified; we therefore do not explicitly write the
appropriate combinations of Dirac matrices between the quark fields. 
The Wilson Operator Product Expansion (OPE) \cite{Wilson} provides the
following expansion for the $T$-product:
\begin{eqnarray}
T\{j(x)j^\dagger(0)\}=C_0(x^2,\mu)\hat 1 +
\sum\limits_n C_n(x^2,\mu) :\hat O_n(x=0,\mu):\ ,
\end{eqnarray}
where $\mu$ is a renormalization scale.
The two-point function (for those cases where only light quarks are
involved) is then expanded in the form 
\begin{eqnarray}
\Pi(p^2)=\Pi_{\rm pert}(p^2,\mu)+\sum_n \frac{C_n}{(p^2)^n}\langle 0| :
\hat O_n(x=0,\mu): |0 \rangle. 
\end{eqnarray}
The QCD vacuum is nonperturbative and its properties are characterized by
the {condensates} -- nonzero vacuum expectation values of gauge-invariant
operators, depending, in general, on the renormalization scale $\mu$, 
$\langle 0 | :\hat O(0,\mu):| 0\rangle\ne 0$
\cite{SVZ, NSVZ1984,NSVZ,Ioffe}. 
Hereafter, we denote $\langle 0|...|0\rangle\equiv \langle ...\rangle$.
In the OPE context, perturbative diagrams describe the contribution of
the unit operator.  
The Wilson coefficients of the local operators $:\hat O(0,\mu):$ are
obtained from perturbative diagrams 
according to known rules \cite{NSVZ1984}. The appropriate diagrams will be
displayed below; we show only diagrams 
containing quark and gluon lines but do not show those diagrams that
contain Faddeev-Popov ghosts. 

It proved efficient to generalize QCD, based on the color gauge group
SU(3), to the case of the color gauge group SU$(N_c)$ and to consider 
the $1/N_c$ expansion of the Green functions in QCD$(N_c)$ in the
so-called 't Hooft limit, the strong coupling constant scaling as
$\alpha_s\sim 1/N_c$ \cite{tHooft}.
For the discussion of exotic states of {\it any} structure [in QCD$(N_c)$
one may have a very rich structure of colorless hadron states] 
the following features of large-$N_c$ QCD are of special significance:
as shown by Witten \cite{Witten}, large-$N_c$ Green functions 
are saturated by noninteracting ordinary mesons. This means, in
particular, that any possible exotic states may appear only 
in $N_c$-subleading 
contributions to the QCD$(N_c)$ Green functions \cite{Witten,Coleman}. 
This is a distinguishing property of exotic tetraquark mesons compared
to ordinary quark-antiquark mesons: the latter appear already in the
$N_c$-leading parts of the QCD$(N_c)$ Green functions. 

The possibility of the existence of narrow exotic states in QCD at large
$N_c$ has been studied in 
recent years in a number of publications \cite{Pelaez,Weinberg,Guo,Knecht,
Cohen1,Cohen2,Maiani1,lms1,lms2,Maiani2,Maiani3,lms3}. 
The present paper makes use of large-$N_c$ arguments in a slightly
different context: We study two-point functions of bilinear quark
currents and of tetraquark currents and discuss the appearance of the
tetraquark bound states in the latter, giving further 
theoretical arguments in the derivation of the tetraquark-adequate
($T$-adequate) sum rules, formulated in our recent publications 
\cite{lms_sr1,lms_sr2}. Here, we make the following new steps: 
\begin{itemize}
\item[(i)] 
We show that the $T$-adequate sum rules in SU$(N_c)$ should be based
on appropriate nonfactorizable parts of the OPE for two-point 
functions of the tetraquark currents. In this way, the tetraquark
contributions are compatible with the well-known 
rigorous property of QCD at large $N_c$: The $N_c$-leading Green
functions are saturated by the ordinary mesons; any 
exotic states may appear only in $N_c$-subleading contributions. 
\item[(ii)] 
We discuss nonperturbative effects in two-point functions of tetraquark
currents [our analyses \cite{lms_sr1,lms_sr2} considered perturbative
diagrams and did not address nonperturbative effects]
and identify those condensates contributions that appear in the
$T$-adequate duality relations and $T$-adequate sum rules. 
\end{itemize}
The paper is organized as follows: 
Section \ref{Sect:M} recalls the large-$N_c$ behavior of the two-point
Green functions of bilinear quark currents 
and compares the OPE with the hadron saturation of these Green functions.
Section \ref{Sect:T} studies the OPE for the two-point function of
tetraquark currents including nonperturbative condensate contributions
and discusses the quark-hadron duality relations that may involve 
possible tetraquark states. 
Section \ref{Sect:C} presents our conclusions and outlook. 

\section{Two-point function of bilinear quark currents
\label{Sect:M}}
Let us start with some uselful algebraic relations for the group
SU$(N_c)$ \cite{Haber}. 
The generators $T^A$, $A=1,\dots,N_c^2-1$, considered in the fundamental
representation, satisfy the color Fierz rearrangement 
\begin{eqnarray}
\label{ColorFierz}
(T^A)_{ii'}(T^A)_{jj'}=\frac12\delta_{ij'}\delta_{i'j}-
\frac1{2N_c}\delta_{ii'}\delta_{jj'},  
\end{eqnarray} 
where the generators are normalized according to 
\begin{eqnarray}
\label{ColorNorm}
{\rm Tr}(T^A T^B)=\frac12 \delta^{AB}. 
\end{eqnarray} 
The relation (\ref{ColorFierz}) suggests that, with respect to counting
an overall $N_c$-leading color 
factor of a Feynman diagram, any gluon line may be replaced by a
$\bar qq$ double line. 
To calculate the $N_c$-subleading terms in the expansion of a Green
function, 
one has to take into account, in the gluon lines, the second term of the
right-hand side of Eq. (\ref{ColorFierz}).  

We now briefly recall the properties of the $1/N_c$-expansion of the
OPE series for bilinear quark currents, $j=\bar qq$. 
\subsection{Perturbative diagrams}
\begin{figure}[!ht]
\centering
\includegraphics[height=1.8cm]{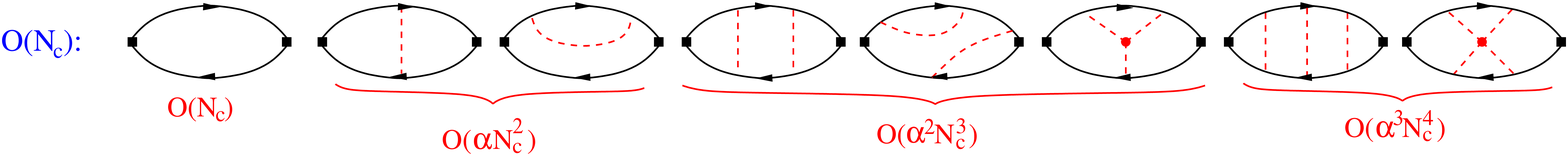}\\
(a)\\
\includegraphics[height=1.7cm]{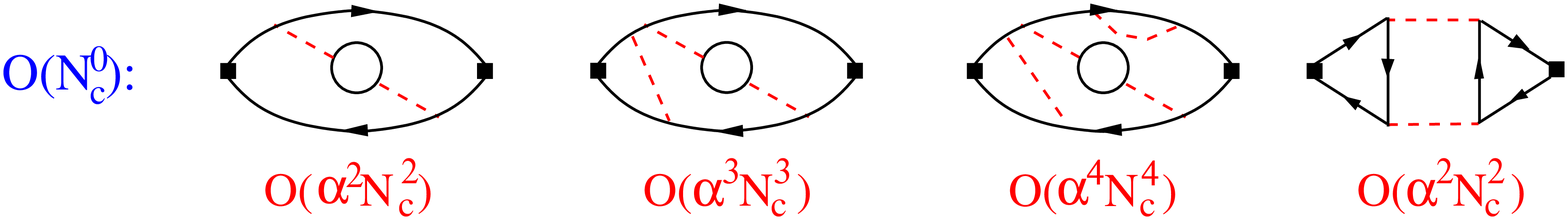}\\
(b)\\
\includegraphics[height=1.9cm]{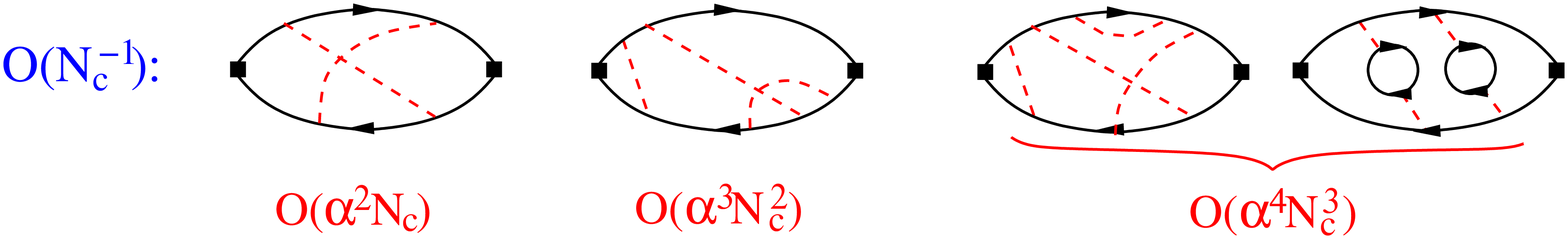}\\
(c)
\caption{Perturbative series for the two-point function of bilinear quark
currents and its classification in powers of $1/N_c$.
(a) $O(N_c)$. $N_c$-leading diagrams are diagrams of planar topology
without sea-quark loops (i.e., containing only the 
loop of valence quarks that enter the 
interpolating current) and with an arbitrary number of planar gluon
exchanges. No annihilation-type diagrams appear at this $N_c$-order. 
Taking into account that $\alpha_s\sim 1/N_c$, they have the overall
dependence $O(N_c)$. 
(b) $O(N_c^0)$. Diagrams with the $N_c$-leading contribution of this
order are shown; these diagrams are (i) 
planar diagrams with one sea-quark loop and an arbitrary number of gluon
exchanges or (ii) quark-annihilation diagrams. 
(c) $O(N_c^{-1})$. Diagrams with $N_c$-leading behavior of this order
are of two different classes: (i) nonplanar diagrams with one gluonic 
handle and an arbitrary number of planar gluon exchanges 
but no sea-quark loops and (ii) planar diagrams containing two sea-quark
loops and an arbitrary number of planar gluon exchanges.}
\label{Figope1}       
\end{figure}

Figure \ref{Figope1} shows diagrams according to their behavior in the
framework of the $1/N_c$ expansion, 
assuming that the strong coupling constant scales as $\alpha_s\sim 1/N_c$. 
When calculating the $1/N_c$ behavior of a diagram, we replace the gluon
line by a double $\bar qq$ line. 
Doing so, we pick up the leading behavior at large $N_c$, but omit
corrections of the order $1/N_c^2$. [The second 
term in Eq.~(\ref{ColorFierz}) contains a factor $1/N_c$ and, in addition,
the number of color loops generated by the second term is reduced 
by one compared to the first term, thus yielding an overall suppression
factor $1/N_c^2$]. So, all diagrams 
with a specific large-$N_c$ behavior generate also contributions to 
lower orders of the $1/N_c$ expansion. For instance, 
the $O(N_c)$ diagrams in Fig.~\ref{Figope1}(a) also generate
contributions to diagrams of the order $O(N_c^{-1})$.
The $N_c$-leading perturbative diagrams of Fig.~\ref{Figope1}(a) are
planar diagrams without sea-quark loops. 
Using the language of intermediate states, these diagrams can be
identified as those 
diagrams that have intermediate valence $\bar qq$ states plus an
arbitrary number of gluons; 
the $N_c$-leading perturbative QCD diagrams do not have cuts
corresponding to four quarks and an arbitrary number of gluons, 
six quarks and an arbitrary number of gluons, etc. Diagrams with
multiquark intermediate states have an $N_c$-subleading behavior. 

\subsection{Power corrections}
\begin{figure}[!h]
\centering
\includegraphics[height=2.0cm]{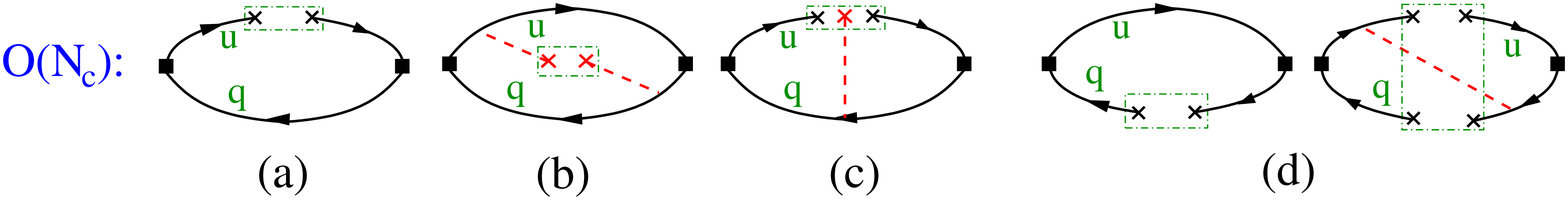}
\caption{Nonperturbative power corrections in $\langle jj^{\dag} \rangle$
of bilinear quark currents:
(a) contribution of dimension-3 quark condensate $\langle \bar uu\rangle$; 
(b) contribution of dimension-4 gluon condensate
$\langle \alpha_sGG\rangle$;
(c) contribution of dimension-5 mixed condensate
$\langle \bar u \sigma_{\mu\nu}G^{\mu\nu}u\rangle$, with
$\sigma_{\mu\nu}=\frac{1}{2i}[\gamma_{\mu},\gamma_{\nu}]$, the $\gamma$s
being the Dirac matrices;
(d) contributions emerging if $q$ is the light quark: 
that of dimension-3 quark condensate $\langle \bar qq\rangle$ 
and that of dimension-6 four-quark condensate
$\langle \bar uu \bar qq\rangle$.}
\label{Figope2}       
\end{figure}

Power corrections are shown in Fig.~\ref{Figope2}: the Wilson
coefficients describing the contribution of the appropriate operators 
may be obtained from the perturbative diagrams of Fig.~\ref{Figope1} by
breaking one or more quark and gluon lines. 
For instance, the diagram of Fig.~\ref{Figope2}(a) provides the Wilson
coefficient of the operator $\bar qq$; 
the diagram of Fig.~\ref{Figope2}(b) gives the Wilson coefficient of
the operator $GG$, where $G$ is the gluon field strength, etc. 

Power corrections scale with $N_c$ as follows: 
\begin{eqnarray}
\langle \bar qq \rangle\sim N_c; \quad \langle \alpha_s GG \rangle
\sim N_c; \quad \langle \bar qq \bar qq\rangle\sim N_c^2.
\end{eqnarray}
Obviously, nonperturbative effects described by the condensates
contribute on the same footing as perturbative effects at each order
in $1/N_c$. 
So, even at $N_c$-leading order QCD is not fully perturbative: although
the strong coupling may be made arbitrary small, 
nonperturbative effects described by the condensates do not disappear and
survive the large-$N_c$ limit. 
\newpage
\subsection{Hadron saturation of two-point function and sum rules}
As is well known \cite{Witten,Coleman}, the spectrum of states of
QCD$(N_c)$ in the limit $N_c\to\infty$ contains towers of an infinite
number of free, stable and noninteracting mesons. Meson-meson elastic
scattering amplitudes are of order $1/N_c^{}$ and the decay
amplitudes of mesons into two mesons are of order $1/N_c^{1/2}$.
The notion of valence quarks takes in the above limit a precise
meaning. Mesons are made of pure $\bar qq$ states, rather than
of $\bar q\bar qqq$ states, which only appear at subleading
orders of $N_c^{}$. Conversely, states, whose $N_c$-leading element
is composed of $\bar q\bar qqq$ states, correspond to
two-meson states. These properties allow us to make a systematic
correspondence between the OPE and the hadron saturation of two-point
correlation functions.

\begin{figure}[!h]
\centering
\includegraphics[height=3.5cm]{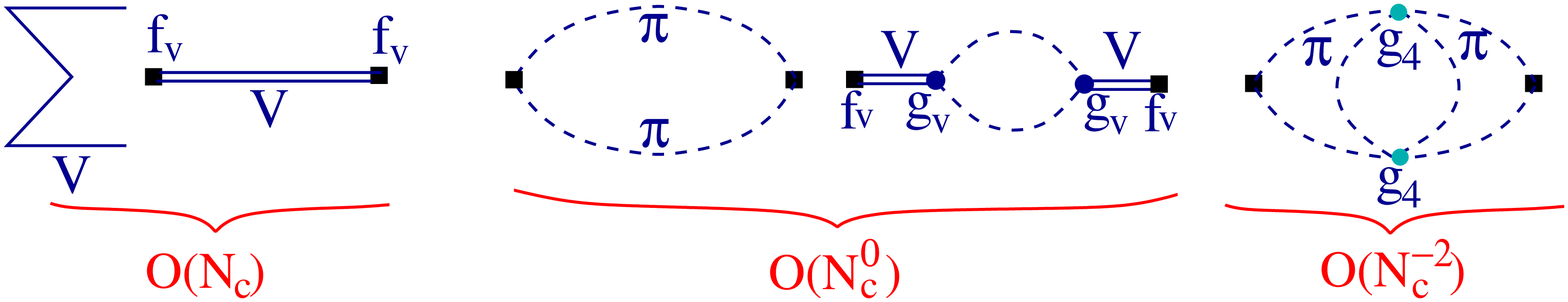}\\
(a)\\
$ $\\
\includegraphics[height=1.5cm]{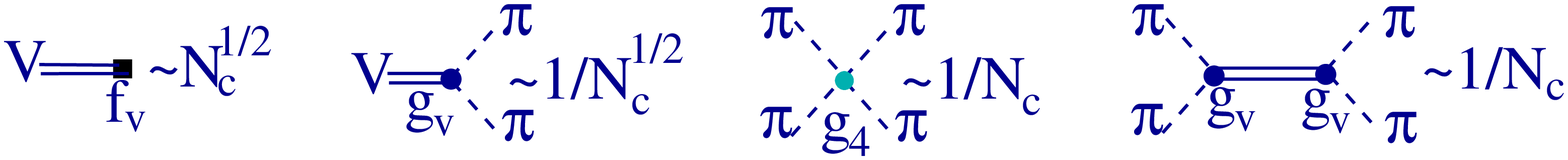}\\
(b)\\
$ $\\
\includegraphics[height=4.0cm]{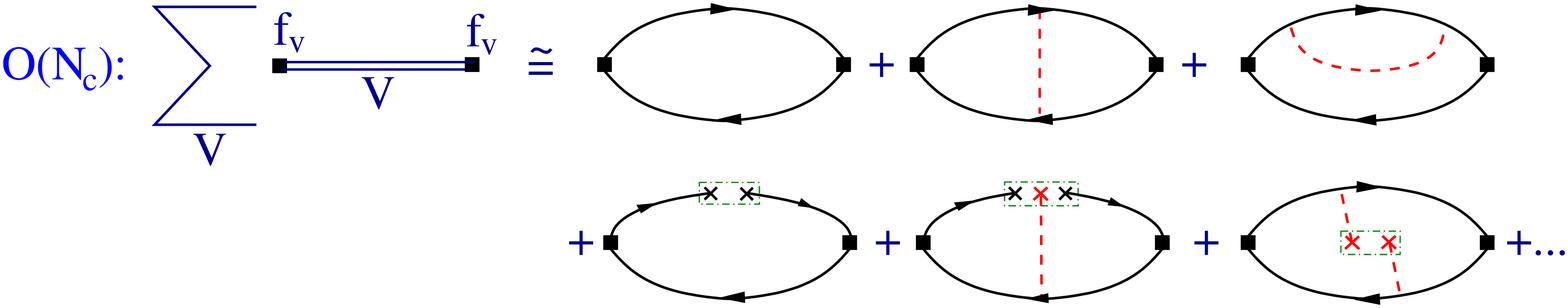}\\
(c)
\caption{Duality relation for the elastic vector two-point function at
large $N_c$: (a) Typical hadron diagrams emerging in the hadron
representation of $\Pi_V$. 
(b) Scaling of the hadron couplings at large $N_c$. 
(c) $N_c$-leading sum rule [of order $O(N_c)$].}
\label{Figope2hadr}       
\end{figure}

Let us consider the two-point function of the elastic vector current,
$V_\mu=\bar q \gamma_\mu q$, and denote it as 
$\Pi^V_{\mu\nu}(x)\equiv \langle T\{ V_\mu(x)V_\nu(0)\}\rangle$.
Obviously, we have light pseudoscalar mesons (hereafter referred to as
pions). Figures~\ref{Figope2hadr}(a,b) show the scaling of 
the hadron diagrams and vertices at large $N_c$ and
Fig.~\ref{Figope2hadr}(c) shows the 
$N_c$-leading sum rule: the sum over stable vector mesons is dual to
the $N_c$-leading OPE. 
Obviously, one can include some $N_c$-subleading effects on the hadron
and/or on 
the OPE side of this sum rule. However, for the consistency of the full
approach, it is mandatory that the $N_c$-leading contribution on the
hadron side matches the $N_c$-leading contribution on the OPE side.
In fact, the vector sum rule in real QCD \cite{SVZ} looks very similar
to the relation shown 
in Fig.~\ref{Figope2hadr}(c) and thus perfectly satisfies the
large-$N_c$ consistency. 

Closing this discussion, we notice that the correlation function of
bilinear quark currents describes the ``minimal'' 
colorless cluster and thus does not contain inside it any other
factorizable colorless clusters. 
As a result, the hadron saturation of the $N_c$-leading part of the
correlation function contains all intermediate hadron states 
with the appropriate quantum numbers, starting from the one-meson state. 


\newpage
\section{OPE for correlators of tetraquark currents
\label{Sect:T}}
Let us now turn to two-point functions of tetraquark currents. We consider
four-quark currents consisting of 
two antiquarks of generic flavors $b$ and $c$ and two quarks of generic
flavors $u$ and
$d$. For the sake of argument, we make two assumptions: First, we assume
that all quark flavors are different -- this simplifies the topology of 
the appropriate QCD diagrams, avoiding, in particular, the discussion of
quark annihilation diagrams.
Our second assumption is that the two antiquarks, $\bar b$ and $\bar c$, 
are heavy and therefore do not produce quark condensates; 
the two quarks, $u$ and $d$, are light and therefore develop local
vacuum condensates (quark condensates 
$\langle \bar uu\rangle$, $\langle\bar dd\rangle$, mixed quark-gluon
condensates $\langle\bar u\sigma_{\mu\nu}T^A G^{A}_{\mu\nu}u\rangle$, 
$\langle \bar d\sigma_{\mu\nu}T^A G^{A}_{\mu\nu}d\rangle$, four-quark
condensates $\langle \bar uu \bar dd \rangle$, etc.). 
These assumptions simplify the discussion but do not change any essential
qualitative feature of our analysis.  

As follows from the property of cluster reducibility of multiquark
operators \cite{lms_sr3}, 
any gauge-invariant multiquark operator can be reduced to a combination
of products of colorless clusters. 
In our case of the $\bar b\bar c ud$ flavor content of the tetraquark
current, colorless clusters of two different 
flavor structures emerge in QCD (see \cite{Jaffe1,Jaffe2}):
\begin{equation}
\label{theta}
\theta_{\bar b u\bar c d}=j_{\bar b u}j_{\bar c d},\ \ \ \ \ \ \ \ 
\theta_{\bar b d\bar c u}=j_{\bar b d}j_{\bar c u},
\end{equation}
with $j_{\bar af}=\bar q_a q_f$.
We therefore should distinguish between the diagrams where quark flavors
in the initial and
final states are combined in the same way (direct diagrams) and 
in a different way (quark-exchange or recombination diagrams).
The Feynman diagrams for the corresponding four-point functions
have different topologies and structures of their four-quark
singularities and therefore require separate analyses
\cite{lms_sr1,lms_sr2}. 
Here, we discuss the direct Green function   
\begin{eqnarray}
\label{Pidir}
\Pi^{\mathrm{dir}}(x)\equiv \langle T\{\theta_{\bar bu\bar cd}(x)
\theta^\dagger_{\bar bu\bar cd}(0)\}\rangle.
\end{eqnarray}
\subsection{Perturbative diagrams}
\begin{figure}[!h]
\centering
\includegraphics[height=2.2cm]{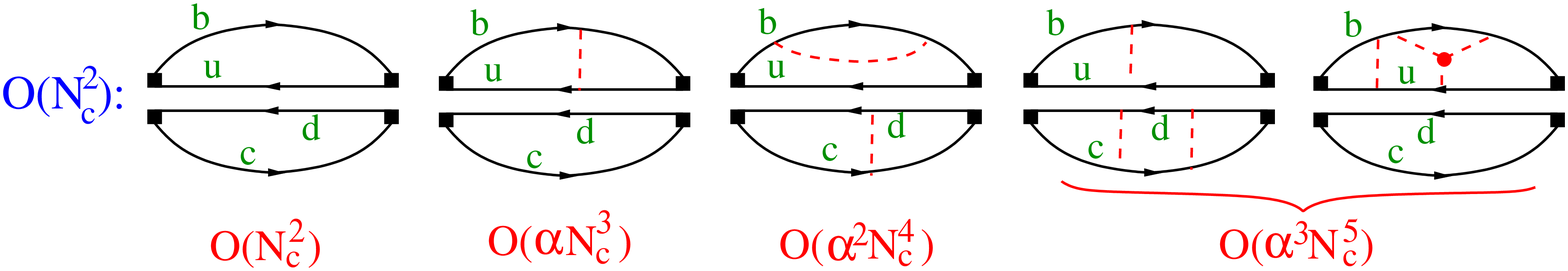}\\
(a)\\
\includegraphics[height=4.8cm]{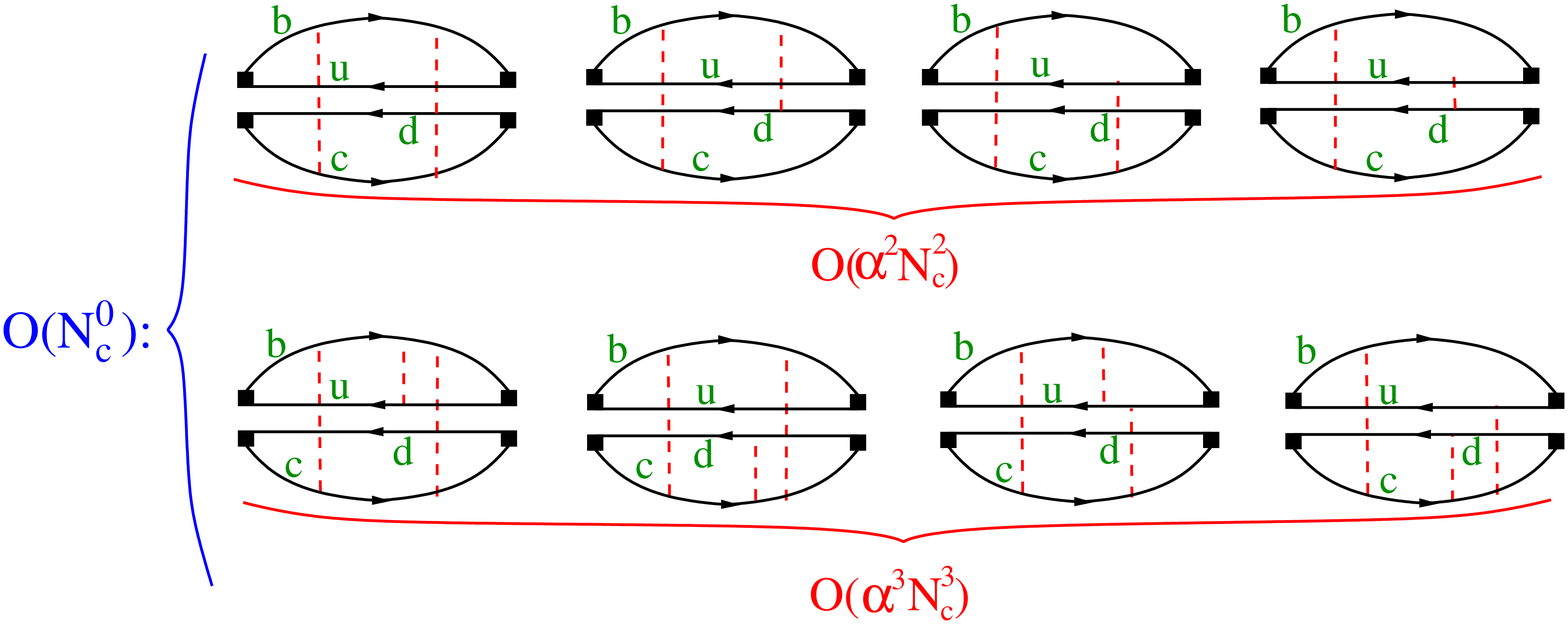}\\
(b)\\
\caption{Typical perturbative diagrams emerging in the OPE for
$\Pi^{\mathrm{dir}}$:
(a) Planar diagrams with an arbitrary number of planar gluon exchanges
inside the $bu$ or $cd$ quark loops, whereas there are no 
gluon exchanges between the quark loops. These diagrams have a
factorizable structure, i.e., may be represented, in coordinate space, 
as a product of two expressions --  one corresponding to the $bu$ loop,
and the other corresponding to the $cd$ loop. They have the 
behavior $\sim N_c^2$. 
(b) Diagrams with two-gluon exchanges between the loops $bu$ and $cd$;
the two gluons may be attached to quarks/antiquarks 
in different quark loops in any combinations. Such diagrams have a
cylinder topology and behave as $N_c^0$ at large $N_c$. 
Adding an arbitrary number of planar gluon exchanges inside the loops
$bu$ or $cd$ does not change the large-$N_c$ scaling behavior.}
\label{Figope3}       
\end{figure}

Figure \ref{Figope3} shows perturbative diagrams in the OPE for
$\Pi^{\mathrm{dir}}$ with different types of gluon exchanges. 

Perturbative diagrams without gluon exchanges between the quark loops
have a factorizable structure (i.e., factorize,
in coordinate space, into the product of two-point functions of bilinear
quark currents, $\Pi_{\bar bu}(x)\Pi_{\bar cd}(x)$, where 
\begin{equation}
\Pi_{\bar bu}(x)\equiv \langle T\{ j_{\bar bu}(x)
j^\dagger_{\bar bu}(0)\}\rangle,\ \ \ \ \ \ \ \  
\Pi_{\bar cd}(x)\equiv \langle T\{ j_{\bar cd}(x)
j^\dagger_{\bar cd}(0)\} \rangle. 
\end{equation}
One can refer to these diagrams as to ``disconnected'' diagrams, bearing
in mind, however,  
that the vertices corresponding to the initial (final) bilinear currents
are connected to each other.\footnote{Notice that the terms
``connected'' and ``disconnected'' in lattice calculations have quite
different meanings, see, e.g., \cite{Guo}.}
So the term ``disconnected'' should apply to the internal parts of the
diagrams. We will prefer to term these diagrams
``factorizable''. Since any of the quark loops behaves as $\sim N_c$,
the factorizable diagrams are of order $N_c^2$. Factorizable diagrams
can be isolated in a unique way and provide the $N_c$-leading behavior
of $\Pi^{\mathrm{dir}}(x)$. 

In the diagrams of Fig.~\ref{Figope3}(b) two quark loops talk to each other
via gluon exchanges. Since both quark loops 
represent colorless clusters, one needs at least two gluons to be
exchanged between the loops. Diagrams with two gluon 
exchanges between the loops and an arbitrary number of planar gluon
exchanges inside each of the loops $bu$ and $cd$ have cylinder topology, 
see Figs.~1 and 2 of Ref.~\cite{lms3}. Their behavior at large $N_c$
is $N_c^0$.  

One can also have three or more gluon exchanges between the quark loops.
All these diagrams have a topology of a 
cylinder with a number of handles. Each handle reduces the large-$N_c$
behavior by two powers of $N_c$.  

\subsection{Diagrams containing condensates}
Diagrams containing condensates may be obtained from the perturbative
diagrams by breaking the internal quark and gluon lines 
and sending the corresponding particles to vacuum condensates.  

\begin{figure}[!h]
\centering
\includegraphics[height=1.5cm]{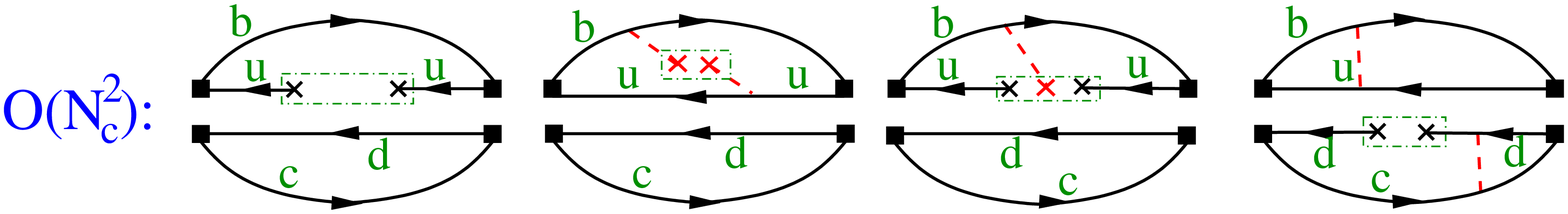}\\
(a)\\
\includegraphics[height=1.5cm]{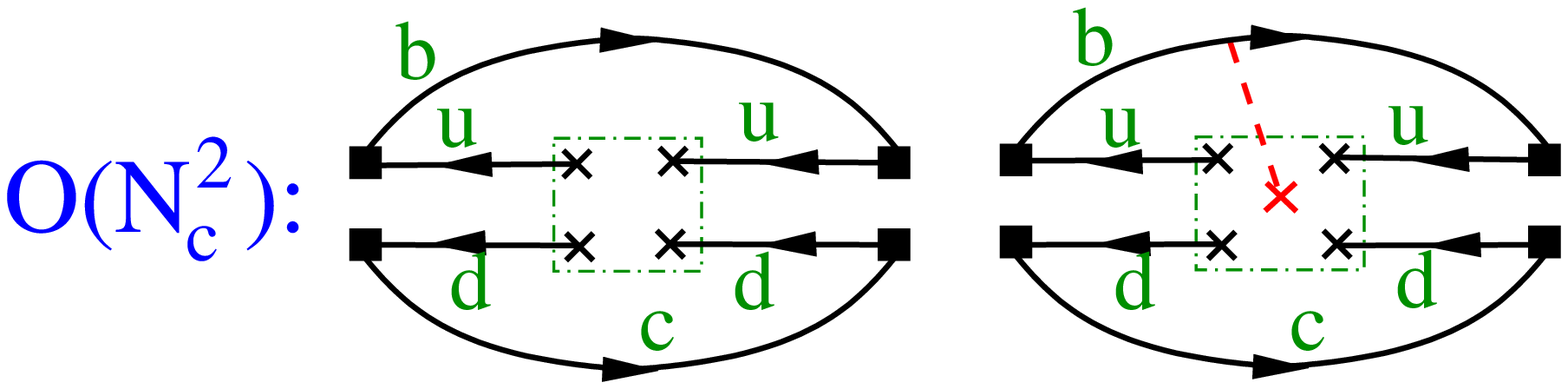}\\
(b)\\
\includegraphics[height=1.5cm]{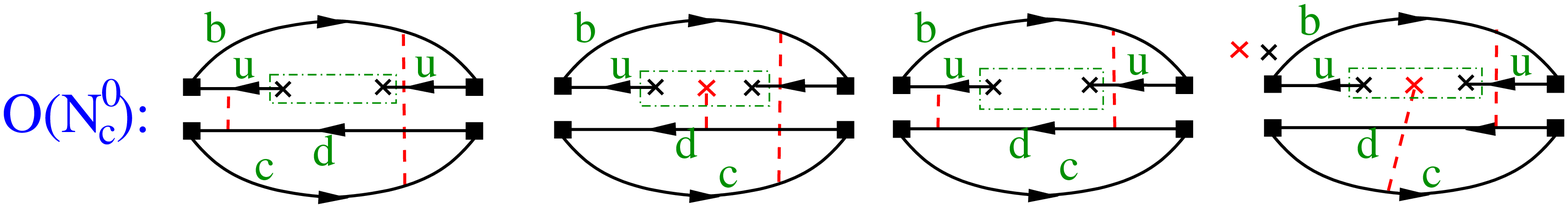}\\
(c)\\
\caption{Typical diagrams belonging to different classes containing
condensate contributions in the OPE for $\Pi^{\mathrm{dir}}$.
(a) Factorizable diagrams obtained by inserting condensate contributions
in perturbative diagrams of Fig.~\ref{Figope3}(a). 
(b) Power corrections of the mixed type: they are proportional to
condensates of dimension-6 or 
higher ($\langle \bar uu\bar dd\rangle$, etc.) and 
obtained by sending to quark or gluon fields from different quark loops
of factorizable perturbative diagrams of Fig.~\ref{Figope3}(a) to vacuum
condensates. 
In the four-quark (and higher) condensates, factorizable and
nonfactorizable parts may be isolated, see Fig.~\ref{Figope5}.  
(c) Power corrections of nonfactorizable type: they are obtained by
sending to the condensate quarks and gluons 
in nonfactorizable diagrams of Fig.~\ref{Figope3}(b).
}
\label{Figope4}       
\end{figure}

Let us start with the factorizable perturbative diagrams of
Fig.~\ref{Figope3}(a). 
By breaking a light-quark or a gluon line in these diagrams, 
one obtains the contributions of condensates of lowest dimensions, the
quark condensate $\langle \bar qq\rangle$, the gluon condensate 
$\langle \alpha_s GG\rangle$, or the mixed quark-gluon condensate,
shown in Fig.~\ref{Figope4}(a).
These diagrams have the same large-$N_c$ behavior, $O(N_c^2)$, as the
original perturbative diagrams of Fig.~\ref{Figope3}(a). Important for
us is that diagrams containing the quark, the gluon, 
and the mixed quark-gluon condensates are of the factorizable type, same
as the original perturbative diagrams.  

A new feature emerges when one calculates the contributions (i.e., the
Wilson coefficients) 
of higher-dimension four-quark and four-quark--gluon condensates, see
Fig.~\ref{Figope4}(b). 

\begin{figure}[!ht]
\centering
\includegraphics[height=1.7cm]{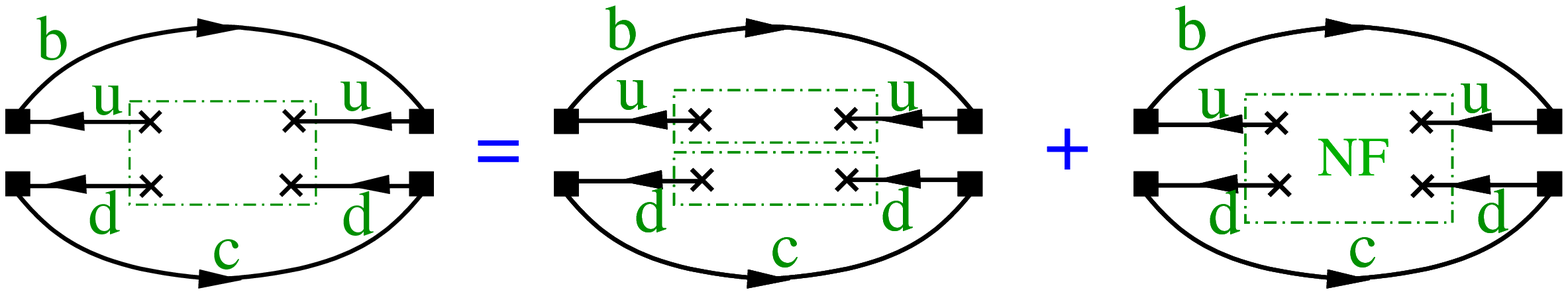}
\caption{Splitting of the condensates of higher dimensions into
factorizable and nonfactorizable (NF) parts. 
The factorizable parts provide the $N_c$-leading contribution, whereas
the nonfactorizable pieces have an
$N_c$-subleading behavior. For instance, $\langle\bar uu\rangle\sim N_c$,
$\langle\bar dd\rangle\sim N_c$,
$\langle\bar uu\bar dd\rangle\sim N_c^2$, whereas
$\langle\bar uu\bar dd\rangle_{\rm NF}\sim N_c^0$.}
\label{Figope5}       
\end{figure}
Here, both light quarks, from the upper and the lower loops, can be sent
to the condensate simultaneously. 
For further analysis, it is convenient to isolate factorizable
contributions from higher-dimension condensates. 
For instance, the four-quark condensate may be split into factorizable
and nonfactorizable (NF) parts in a unique way, see Fig.~\ref{Figope5}: 
\begin{eqnarray}
\langle \bar uu\bar dd \rangle \equiv \langle \bar uu \rangle
\langle \bar dd \rangle + \langle \bar uu\bar dd \rangle_{\rm NF}. 
\end{eqnarray}

The relevance of isolating factorizable parts out of the condensates of
higher dimensions will become clear shortly. 
  
\subsection{Isolating the factorizable part from the OPE for two-point
functions of tetraquark currents $\Pi^{\mathrm{dir}}$}
\begin{figure}[!b]
\centering
\includegraphics[width=17.5cm]{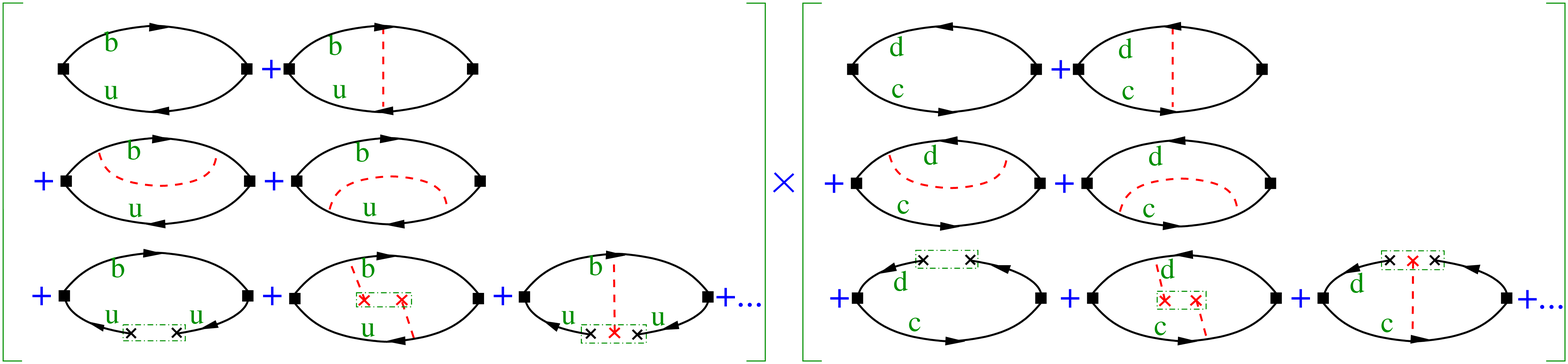}\\
(a)\\
\includegraphics[height=1.6cm]{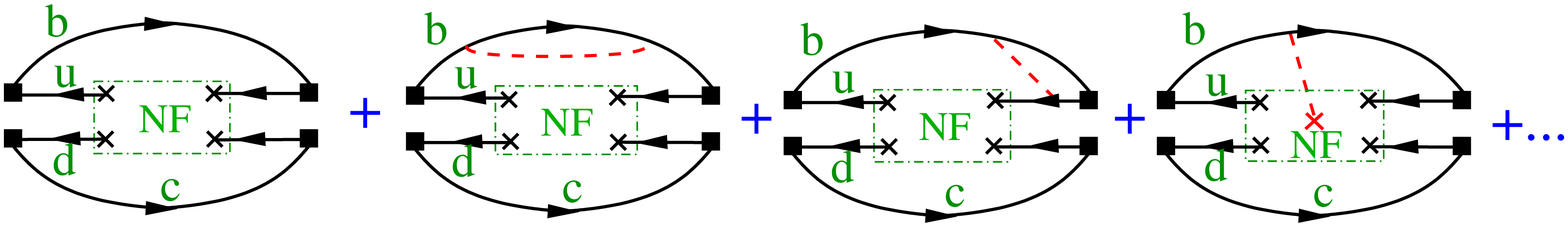}\\
(b)\\
\includegraphics[height=1.6cm]{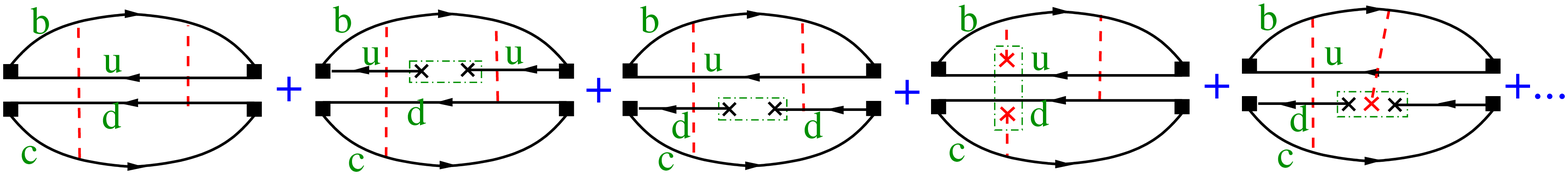}\\
(c)\\
\caption{Different types of contributions to the OPE for
$\Pi^{\mathrm{dir}}$ corresponding to the decomposition (\ref{opedir}): 
(a) Factorizable part of the OPE given by the product of
$\Pi_{\bar bu}(x)$ and $\Pi_{\bar cd}(x)$; 
$\Pi_{\bar bu}(x)$ and $\Pi_{\bar cd}(x)$ here contain both perturbative
and nonperturbative condensate contributions. 
(b) Typical diagrams for the nonfactorizable contribution
$\Pi^{\mathrm{dir}}_{\rm NF,1}(x)$.  
(c) Nonfactorizable contribution $\Pi^{\mathrm{dir}}_{\rm NF,2}(x)$. 
Only those diagrams where one gluon is attached to the quarks $b$ and $c$,
and the other gluon is attached to the quarks $b$ and $d$ are displayed.
Diagrams corresponding to other gluon exchanges between the $bu$ and $cd$ 
quark loops and the appropriate power corrections can be easily drawn. }
\label{Figope6}       
\end{figure}

We are now fully prepared to isolate the factorizable part from the OPE
for the direct two-point function of the tetraquark currents
$\Pi^{\mathrm{dir}}$, including both perturbative and nonperturbative 
condensate contributions. 
At the level of diagrams, one may naively suspect the following
decomposition of the OPE for $\Pi^{\mathrm{dir}}$ shown in
Fig.~\ref{Figope6}:
$$\Pi^{\mathrm{dir}}(x)=\Pi_{\bar bu}(x)\Pi_{\bar cd}(x)
+\Pi^{\mathrm{dir}}_{\rm NF,1}(x)+\Pi^{\mathrm{dir}}_{\rm NF,2}(x).$$
As we shall see, this formula contains double counting of some of the
nonperturbative contributions, and the correct decomposition is a bit
more tricky. 
  
The factorizable part, $\Pi_{\bar bu}(x)\Pi_{\bar cd}(x)$, including
the appropriate nonperturbative condensate contributions is obvious and
is shown in Fig.~\ref{Figope6}(a). 

\begin{figure}[!t]
\centering
\includegraphics[height=2.5cm]{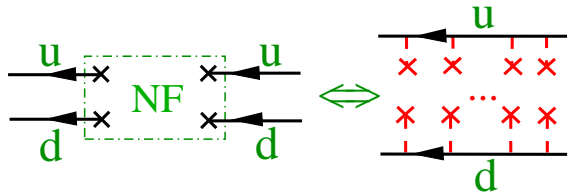}
\caption{The emergence of the nonfactorizable part of the four-quark
condensate $\langle \bar uu \bar dd \rangle$  
through interactions with the nonperturbative gluon background.}
\label{Figope5c}       
\end{figure}

More cumbersome are nonfactorizable (NF) contributions. Here, we
encounter two types of such NF contributions: 
\begin{itemize}
\item
$\Pi^{\mathrm{dir}}_{\rm NF,1}(x)$, shown in Fig.~\ref{Figope6}(b),
describes nonfactorizable parts of the condensates of higher dimension
corresponding to Fig.~\ref{Figope4}(b); somewhat tricky, these NF
nonperturbative power corrections are generated by {\it factorizable}
perturbative diagrams of Fig.~\ref{Figope3}(a). Radiative corrections
due to gluon exchanges inside the loops $bu$ and $cd$ 
(and not between the loops) are included in this class of NF contributions. 

To understand the proper way to take into account such contributions, let
us recall that 
nonzero vacuum condensates emerge due to interactions with the
nonperturbative soft gluon fields. 
If we look into the anatomy of $\langle \bar uu\bar dd\rangle_{\rm NF}$, 
this quantity is nonzero due to the interactions with the soft gluons of
the type shown 
in Fig.~\ref{Figope5c}. Precisely the same nonperturbative corrections
emerge in the 
nonperturbative contributions related to the diagrams of
Fig.~\ref{Figope6}(c) 
where gluons are exchanged between the $u$ quark of the 
$bu$ quark loop and the $d$ quark of the $cd$ quark loop. [This actually
explains the fact that the large-$N_c$ behaviors of 
$\langle \bar qq\rangle^2$ and $\langle \bar qq\bar qq\rangle_{\rm NF}$
differ from each other.]

Taking into account both effects described by
$\Pi^{\mathrm{dir}}_{\rm NF,1}(x)$
and by nonperturbative corrections in those parts of
$\Pi^{\mathrm{dir}}_{\rm NF,2}(x)$
which correspond to two-gluon exchanges between $u$ and $d$ quarks from
the different quark loops would be a double counting. 
We therefore take the appropriate nonperturbative contributions into
account as a part of $\Pi^{\mathrm{dir}}_{\rm NF,2}(x)$ and omit
$\Pi^{\mathrm{dir}}_{\rm NF,1}(x)$. 

\item
$\Pi^{\mathrm{dir}}_{\rm NF,2}(x)$, Fig.~\ref{Figope6}(c), describes
the ``genuinely'' nonfactorizable perturbative diagrams of
Fig.~\ref{Figope3}(b) and 
power corrections of Fig.~\ref{Figope4}(c); the latter are obtained
via the conventional rules by breaking the propagating lines of light
quarks and gluons in the 
perturbative diagrams of Fig.~\ref{Figope3}(b). 
\end{itemize}
In the end, the proper decomposition of $\Pi^{\mathrm{dir}}$ that avoids
the double counting of the nonperturbative corrections has the form 
\begin{eqnarray}
\label{opedir}
\Pi^{\mathrm{dir}}(x)=\Pi_{\bar bu}(x)\Pi_{\bar cd}(x)
+\Pi^{\mathrm{dir}}_{\rm NF,2}(x).
\end{eqnarray}

Let us insert the full system of hadron states in the factorizable part.
We then obtain 
\begin{eqnarray}
\label{3.6}
\Pi_{\bar bu}(x)\equiv \langle T\{ j_{\bar bu}(x)
j^\dagger_{\bar bu}(0)\} \rangle = \sum_{h_{\bar bu}}R_{\bar bu}(x),\\
\label{3.7}
\Pi_{\bar cd}(x)\equiv \langle T\{ j_{\bar cd}(x)
j^\dagger_{\bar cd}(0)\} \rangle = \sum_{h_{\bar cd}}R_{\bar cd}(x),
\end{eqnarray}
where $R_{\bar bu}(x)$ and $R_{\bar cd}(x)$ are the quantities coming
from hadron saturation, the explicit form of which is irrelevant.
Important for us is the fact that the sum runs over the full system of
hadron states 
with flavors $\bar bu$ ($h_{\bar bu}$) and ${\bar cd}$ ($h_{\bar cd}$),
respectively. 
Consequently, the system of the intermediate hadron states that emerges
in the factorizable
part of $\Pi^{\mathrm{dir}}$  is just the direct product of these two
systems, $h_{\bar bu}\otimes h_{\bar cd}$.
No other hadron state, in particular, no exotic state, may contribute here.
So, we conclude that an exotic state, if it exists in the hadron spectrum
of $\bar bu\bar cd$ states,
contributes only to the nonfactorizable part of an exotic correlation
function. 

\newpage
\subsection{$T$-adequate sum rule and the couplings of tetraquark
bound states to tetraquark currents}

The $T$-adequate sum rule emerges after Eqs.~(\ref{3.6}) and (\ref{3.7})
have been taken into account, leading 
to exact cancellations between the factorizable $O(N_c^2)$ contributions
on the OPE side and the 
hadron side of the duality relation for the direct two-point function
$\Pi^{\mathrm{dir}}$. 
\begin{figure}[!ht]
\centering
\includegraphics[width=15cm]{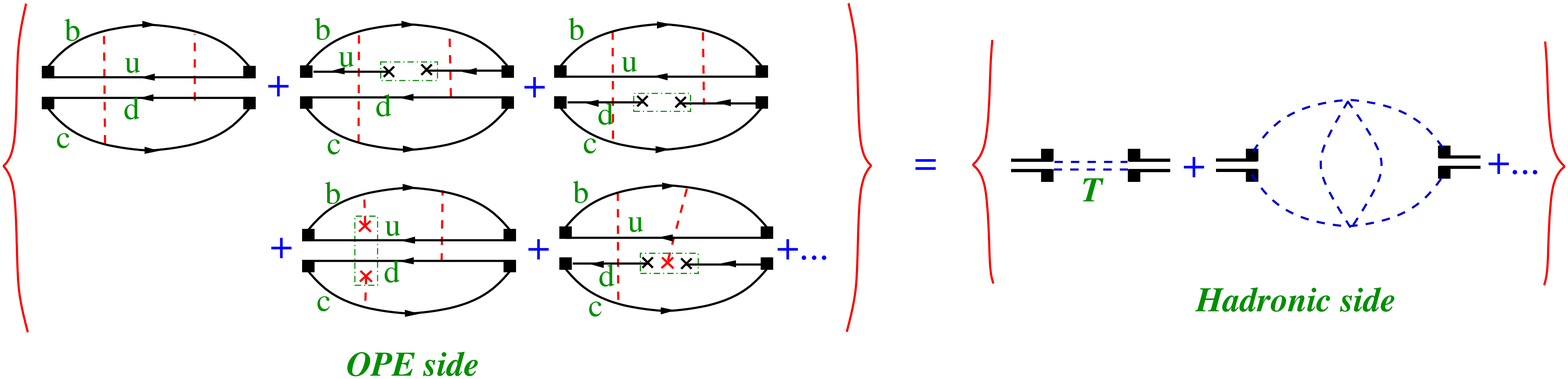}
\caption{$T$-adequate sum rule that emerges after the exact cancellations
of factorizable contributions on the OPE side vs the hadronic side 
have been taken into account. The OPE side contains nonfactorizable
diagrams with two gluon exchanges between the quark loops $bu$ and $cd$.
A typical contribution with gluon exchanges between the loops, joining
$bc$ and $bd$ quark pairs, 
and the corresponding condensate corrections are explicitly shown.
[Noteworthy, all appropriate condensate contributions are 
obtained according to the known rules \cite{NSVZ1984} (i.e., by
breaking the lines of light quarks and gluons) from the perturbative
nonfactorizable Feynman diagrams with two or more gluon exchanges
between the loops; no other condensate contributions appear.]
The dots stand for other two-gluon exchanges (i.e., joining 
$uc$ and $ud$, $bc$ and $ud$, $uc$ and $bd$ pairs from the different
loops). All nonfactorizable diagrams scale as $N_c^0$.
The hadronic side contains the assumed tetraquark contribution and
nonfactorizable meson interaction diagrams.}
\label{Figope7}       
\end{figure}
Figure \ref{Figope7} shows the corresponding $T$-adequate sum rule: its
OPE side contains nonfactorizable diagrams of order $O(N_c^0)$ 
(both perturbative and condensate contributions); its hadron side
contains the assumed tetraquark contribution and nonfactorizable 
meson diagrams. The contribution of the tetraquark $T$ of flavor
content $\bar b\bar c ud$ and mass $M$ to the hadronic side in
momentum space 
has the form\footnote{To be more precise, there are two tetraquark
currents of global  $\bar b\bar c ud$ flavor content:
$\theta_{\bar b u \bar cd}$ and $\theta_{\bar b d \bar c u}$
[Eq.~(\ref{theta})]. 
Respectively, there might be two tetraquark states of the
$\bar b\bar c ud$ flavor content, 
$T_1$ and $T_2$: $T_1$ couples stronger to the $\theta_{\bar bu \bar cd}$
current (the corresponding coupling scales as $N_c^0$) 
and weaker to the $\theta_{\bar bd \bar cu}$ current (the corresponding
coupling scales as $1/N_c$), and vice versa for $T_2$. So, the
contribution of $T_1$ to the correlation function (\ref{Pidir}) scales
like $N_c^0$, whereas the contribution of $T_2$ is suppressed and
scales like $1/N_c^2$. 
These subtleties are, however, a bit outside the main discussion of this
paper, so we refer for details to Sect. 2 of \cite{lms3}.} 
\begin{eqnarray}
f^2_T\frac{1}{M^2-p^2}
\end{eqnarray}
and is expressed via the tetraquark coupling to the interpolating tetraquark
current: 
\begin{eqnarray}
f_T=\langle 0|\theta_{\bar bu \bar cd}|T\rangle. 
\end{eqnarray}
Obviously, the $T$-adequate sum rule implies $f_T\sim N_c^0$. 
This feature is in full agreement with the known property of large-$N_c$
QCD that only noninteracting ordinary mesons saturate the $N_c$-leading
QCD diagrams. 
We emphasize once more that for the consideration of exotic states,
the factorizable part of the OPE is irrelevant. 

Moreover, we would like to point out the following qualitative difference
between the correlation functions of tetraquark versus those of bilinear
quark currents: As we have discussed, the existence of stable vector
mesons in the hadron spectrum at large $N_c$ is required by matching the
large-$N_c$ behavior of 
the OPE side and of the hadron side of the vector two-point function
$\Pi^V$; without vector mesons populating the $O(N_c)$ part of the hadronic
side no matching may be obtained. 
For the two-point functions of the tetraquark currents, the situation is
qualitatively different: 
the factorizable OPE and hadronic sides match each other just due to the
duality relations for the two-point functions of the bilinear currents. 
The nonfactorizable OPE side and its nonfactorizable hadronic side have
the same large-$N_c$ behavior with or without the tetraquark bound state. 
So, the existence of narrow tetraquark hadrons cannot be established
merely on the basis of the large-$N_c$ behavior of the exotic Green
functions; large-$N_c$ QCD does not exclude narrow exotic states in
$N_c$-subleading parts of the Green functions, but remains consistent
also if such exotic states do not exist in the hadron spectrum.

\section{Conclusions and Outlook \label{Sect:C}}
We discussed in great detail the OPE for two-point Green functions of
the bilinear and quadrilinear colorless quark currents
at large $N_c$ and emphasized the qualitative differences between these
two objects: 
\begin{itemize}
\item[(i)] In the case of two-point functions of bilinear quark currents, 
the contributions of single-meson states with appropriate quantum
numbers emerge in the $N_c$-leading part of the Green function. 
Matching the large-$N_c$ behavior of the OPE series and of the
hadron saturation series requires the existence of stable mesons with
large couplings, $f_V\sim \sqrt{N_c}$, in the limit $N_c\to\infty$.
The typical QCD sum rule then relates the $N_c$-leading $O(N_c)$ part
of the OPE to the $O(N_c)$ part of the sum over hadron states, and is
therefore fully consistent at large $N_c$.  
\item[(ii)]
In the case of two-point functions of tetraquark currents, the 
$N_c$-leading part of the OPE factorizes into a product of two
colorless clusters. Each of them is saturated by the 
ordinary hadrons that may emerge in the quark-antiquark correlation
functions with appropriate quark-flavor content. 
As a result, tetraquark states (whatever generalization of the $N_c=3$
tetraquark to $N_c\ne 3$ is considered) 
cannot contribute to the $N_c$-leading parts of the Green functions. 
This property fully agrees with the well-known rigorous property of
large-$N_c$ QCD: $N_c$-leading Green functions are fully saturated by
noninteracting ordinary mesons. The contribution of any colorless state
with a more complicated quark structure, for instance, of an exotic meson,
may only appear in $N_c$-subleading nonfactorizable parts of the Green
functions of tetraquark currents. Moreover, this property is perfectly
satisfied by the $T$-adequate QCD sum rules formulated in 
\cite{lms_sr1,lms_sr2}: one of the outcomes of the $T$-adequate sum
rules is the scaling of the tetraquark coupling to the tetraquark current 
$f_T\sim N_c^0$.
In the present paper, we have complemented our previous analysis with
a detailed discussion of the vacuum condensate contributions.   
\end{itemize}

Existing typical applications of QCD sum rules to the analysis of the
tetraquark candidates (see, e.g., recent publications
\cite{pimikov,wang,narison})
relate the tetraquark properties to the low-energy part of the
factorizable two-point Green functions of the tetraquark currents.
Such an approach copies the route of the sum-rule analysis of the
ordinary mesons and does not take into account the fundamental
differences between the correlation functions of the bilinear quark
currents and of the tetraquark currents. The tetraquark properties are then
extracted exclusively from those parts
of QCD Green functions which do not have tetraquarks as intermediate
states \cite{pimikov,wang,narison}, a feature which  
does not seem to us physically meaningful.
Also, one can easily take the large-$N_c$ limit of the corresponding
sum-rule analytic expressions for the couplings to immediately find
that $f_T\sim N_c$. 
This would mean that, in contradiction to the rigorous property of QCD
at large $N_c$, tetraquark poles appear in the $N_c$-leading QCD diagrams. 
$T$-adequate sum rules of \cite{lms_sr1,lms_sr2},
now complemented with the appropriate account of condensate contributions,
are free from these shortcomings and lead to fully consistent relations.

\acknowledgments 
D.~M. and H.~S. are grateful for support under joint CNRS/RFBR Grant
No. PRC Russia/19-52-15022. H.~S. acknowledges support from the EU
research and innovation program Horizon 2020, under Grant agreement
No. 824093.
D.~M. would like to thank the Organizers of the MIAPP program
``Deciphering Strong-Interaction Phenomenology through Precision
Hadron-Spectroscopy'' held 7-31 October 2019 at the Excellence Cluster
``Universe'' in Garching, Germany, for financial support of his
participation at this workshop, where a part of this work was presented.


\end{document}